\documentclass[twocolumn,letter]{jpsj3}
\usepackage{txfonts}

\usepackage{color}
\usepackage{graphicx}
\usepackage{dcolumn}
\usepackage{bm}

\newcommand{\nc}{\newcommand}
\nc{\be}{\begin{eqnarray}}
\nc{\ee}{\end{eqnarray}}
\nc{\bea}{\begin{eqnarray}}
\nc{\eea}{\end{eqnarray}}
\nc{\bean}{\begin{eqnarray*}}
\nc{\eean}{\end{eqnarray*}}
\nc{\mb}{\mbox}
\nc{\rnc}{\renewcommand}
\nc{\vk}{{\bm k}}
\nc{\vx}{\mb{\bf x}}
\nc{\br}{\mb{\bf r}}
\nc{\bv}{\mb{\bf v}}
\nc{\bp}{\mb{\bf p}}
\nc{\ve}{\mb{\bf e}}
\nc{\vz}{\hat {\mb{\bf z}}}
\nc{\vp}{\mb{\boldmath$p$}}
\nc{\vb}{\mb{\boldmath$b$}}
\nc{\rr}{\mb{\boldmath$r$}}
\nc{\vR}{\mb{\boldmath$R$}}
\nc{\vj}{\mb{\boldmath$j$}}
\nc{\vg}{\mb{\boldmath$g$}}
\nc{\vm}{\mb{\boldmath$m$}}
\nc{\vd}{\mb{\boldmath$d$}}
\nc{\hd}{\mb{\boldmath$\hat{d}$}}
\nc{\vD}{\mb{\boldmath$D$}}
\nc{\vF}{\mb{\boldmath$F$}}
\nc{\vG}{\mb{\boldmath$G$}}
\nc{\vI}{\mb{\boldmath$I$}}
\nc{\vW}{\mb{\boldmath$W$}}
\nc{\x}{\mb{\boldmath$x$}}
\nc{\A}{\mb{\boldmath$A$}}
\nc{\va}{\mb{\boldmath$a$}}
\nc{\vv}{\mb{\boldmath$v$}}
\nc{\vq}{\mb{\boldmath$q$}}
\nc{\vn}{\mb{\boldmath$n$}}
\nc{\vJ}{\mb{\boldmath$J$}}
\nc{\vS}{\mb{\boldmath$S$}}
\nc{\vs}{\mb{\boldmath$\sigma$}}
\nc{\vE}{\mb{\boldmath$E$}}
\nc{\vB}{\mb{\boldmath$B$}}
\nc{\vM}{\mb{\boldmath$M$}}
\nc{\vL}{\mb{\boldmath$L$}}
\nc{\vpsi}{\mb{\boldmath$\psi$}}
\nc{\vphi}{\mb{\boldmath$\varphi$}}
\nc{\Vphi}{\mb{\boldmath$\phi$}}
\nc{\Vomega}{\mb{\boldmath$\Omega$}}
\nc{\ipsi}{\it{\Psi}}
\nc{\vepsilon}{\mb{\boldmath$\epsilon$}}
\nc{\valpha}{\mb{\boldmath$\alpha$}}
\nc{\vgamma}{\mb{\boldmath$\gamma$}}
\nc{\vomega}{\mb{\boldmath$\omega$}}
\nc{\vmu}{\mb{\boldmath$\mu$}}
\nc{\vt}{\mb{\boldmath$\tau$}}
\nc{\vT}{\mb{\boldmath$T$}}
\nc{\vpi}{\mb{\boldmath$\pi$}}
\nc{\nab}{\nabla}
\nc{\ov}{\overline}
\nc{\cdott}{\!\cdot\!}
\nc{\cdottt}{\!\!\cdot\!}
\nc{\LL}{\Big{\langle}}
\nc{\RR}{\Big{\rangle}}
\nc{\LR}{\Bigm{|}}
\nc{\vP}{\mb{\boldmath$P$}}
\nc{\nnn}{\nonumber\\}
\rnc{\figurename}{FIG.}

\title{Weyl semimetal phase in solid-solution narrow-gap semiconductors}

\author{Daichi Kurebayashi and Kentaro Nomura}
\inst{Institute for Materials Research, Tohoku University, Sendai 980-8577, Japan} 

\abst{
We theoretically investigate ferromagnetic ordering in magnetically doped solid-solution narrow-gap semiconductors with the strong spin-orbit interaction such as Cr-doped Bi$_2$(Se$_x$Te$_{1-x}$)$_3$.
We compute the spontaneous magnetization of impurities and itinerant electrons, and estimate the critical temperature as a function of the concentration of magnetic dopants and the strength of the spin-orbit interaction.
It is found that the critical temperature is proportional to the concentration of dopants and enhanced with the strong spin-orbit interaction.
It is also found that the ferromagnetic transition could make the system turn to the Weyl semimetal which possesses a pair of Weyl points separating in the momentum space.
}

\begin{document}
\maketitle



Electric devices generally take advantage of the charge of electrons, whereas magnetic devices are used for recording information involving electron spin. 
The realization of materials that combine semiconducting behavior with robust magnetism has been a central issue of a research field of semiconductor spintronics\cite{Wolf2001,Jungwirth2006}.
It has been established that several compound semiconductors become ferromagnetic when doped with Mn.\cite{Jungwirth2006} 
Magnetic order originates from coupling between magnetic element moments which is mediated by conduction-band electrons or valence-band holes.
Since the Curie temperatures, $T_c$, of these dilute magnetic semiconductors (DMSs) are not as high as those of ferromagnetic metals, a search for new materials is a central issue in this field.

The Weyl semimetals are recently predicted novel magnetic materials with the pseudo-relativistic linear dispersions and magnetic ordering.
Near the band-touching points (Weyl points), the excitations are described by the massless Dirac-Weyl Hamiltonian. 
Quasiparticles, the Weyl fermions, are assigned a chirality, and the bulk band gap opens only if the two Weyl fermions with opposite chirality meet each other.
This topological behavior originates in the nonzero Berry curvature enclosing a Weyl point. 
The Weyl semimetal phase was first proposed to realize in pyrochlore iridates such as A$_2$Ir$_2$O$_7$, where A = Y or a rare earth such as Eu, Nd, and Sm.\cite{Wan2011a}
Spinels based on osmium\cite{Wan2011b}
 and HgCr$_2$Se$_4$\cite{Xu2011}
have also been proposed as candidates.
Another route to engineer Weyl semimetals is the use of heterostructures of topological insulators.\cite{Burkov2011}
Despite several theoretical approaches have been proposed\cite{Wan2011a,Wan2011b,Xu2011,Burkov2011,Weyl2014}, currently there are no clearly established materials with Weyl nodes near the chemical potential.

Other possible candidates for Weyl semimetals with simple setup are Bi$_2$Se$_3$ family doped with transition metals\cite{TPT3}.
Bi$_2$Se$_3$ is known to be a three-dimensional topological insulator (TI)\cite{review_TI}.
In one of those materials, Cr-doped Bi$_2$Se$_3$, a ferromagnetic state is observed by magnetization measurements.\cite{ex_BiSe-mag,J.Zhang2013}
Unlike conventional DMSs where magnetic order is mediated by itinerant electrons, because of strong spin-orbit coupling, ferromagnetic ordering is induced by the Van-Vleck paramagnetism in Bi$_2$Se$_3$ family.\cite{Van-Vleck}

Recently the angle-resolved photoemission spectroscopy (ARPES) demonstrates that, in Cr-doped Bi$_2$(Se$_x$Te$_{1-x}$)$_3$, the bulk band gap becomes smaller as substituting tellurium for selenium and, eventually, closes at the $\Gamma$ point with the selenium concentration $x\approx60\%$ which corresponds to the topological phase transition point.\cite{TPT1,TPT2}
Since the band gap becomes zero with this selenium concentration, it is easy to invert valence and conduction bands by exchange fields\cite{Dirac-Weyl}, and the system turns into the Weyl semimetal phase.

In this work we examine the realization of the Weyl semimetal phase in bulk solid-solution narrow-gap semiconductors doped with magnetic atoms such as Cr and Fe.
As a specific model, we take a hexagonal lattice model with parameters of Bi$_2$Se$_3$ and vary the strength of spin-orbit coupling and the magnetic impurity concentration.
We estimate the critical temperature of the ferromagnetic phase transition by computing the spontaneous magnetization of impurities and itinerant electrons, and propose the diagram of topological phases for Bi$_2$Se$_3$ family.


We introduce an effective lattice model for a narrow gap semiconductor.
For simplicity we consider the highest valence band and the lowest conduction band\cite{BiSe-model}, each band has two-fold degeneracy in the presence of time-reversal symmetry (TRS).
The Hamiltonian is given by
\bea
 H_0&=&\sum_{\vk}c^{\dag}_{\vk}{\cal H}_0({\vk})c^{}_{\vk}
\eea
where $c^{\dag}_{\vk}=(c^{\dag}_{\vk\uparrow -},c^{\dag}_{\vk\downarrow -},c^{\dag}_{\vk\uparrow +},c^{\dag}_{\vk\downarrow +})$ is an electron creation operator with the wave vector $\vk$, $\pm$ is the band index showing parity, and
\bea
\mathcal{H}_0({\vk})=\sum_{i=1}^3 R_i({\vk})\, \alpha_i +  m({\vk})\, \alpha_4 + \varepsilon({\vk})I
\eea
discribes the band structure of the system. ${\boldsymbol \alpha}=(\alpha_1,\alpha_2,\alpha_3,\alpha_4)$ where
\bea
 \alpha_{i}=
\begin{pmatrix}
0 & \sigma_i \\
\sigma_i & 0
\end{pmatrix},
\ \ 
\alpha_4=
\begin{pmatrix}
I & 0 \\
0 & -I
\end{pmatrix},
\eea
with $\sigma_i$ being Pauli matrices, and $R_i({\vk})$, $m({\vk})$, and $\varepsilon({\vk})$ are taken to be the hexagonal lattice,
\bea
R_1({\vk}) &=& \frac{2}{\sqrt{3}}A_1\sin\left(\frac{\sqrt{3}}{2}k_x\right)\cos\left(\frac{1}{2}k_y\right),\\
R_2({\vk}) &=& \frac{2}{3}A_1\left[\cos\left(\frac{\sqrt{3}}{2}k_x\right)\sin\left(\frac{1}{2}k_y\right)+\sin\left(k_y\right)\right],\\
R_3({\vk}) &=& A_2\sin(k_z),
\eea
\bea
\hspace{-3mm}\nonumber m({\vk})  &=&m_0-B_2\big[2-2\cos(k_z)\big]\\
&&\hspace{-10mm}-\frac{4}{3}B_1\left[3-2\cos\left(\frac{\sqrt{3}}{2}k_x\right)\cos\left(\frac{1}{2}k_y\right)-\cos\left(k_y\right)\right], \\
\hspace{-3mm}\nonumber \varepsilon({\vk})&=&-\epsilon_F+D_2\big[2-2\cos(k_z)\big]\\
&&\hspace{-10mm}+\frac{4}{3}D_1\left[3-2\cos\left(\frac{\sqrt{3}}{2}k_x\right)\cos\left(\frac{1}{2}k_y\right)-\cos\left(k_y\right)\right].
\eea
$m_0$ corresponds to the band gap at the Gamma point, which is related to the strength of the spin-orbit interaction. $A_1,A_2,B_1,B_2,D_1,D_2$ and $\epsilon_F$ are material dependent parameters determined to fit the $ab$ $initio$ calculation.\cite{BiSe-model,BiSe-model2}

Next we introduce the magnetic interaction between electrons and localized magnetic impurities:
\be
H_J=J\sum_{I=1}^{N_S}{\bm S}({\bm r}_I)\cdot c_I^\dagger{\bm\Sigma} c_I,
\ee
where ${\bm r}_I$ is the position of the magnetic impurity and ${\bm S}({\bm r})$ is its magnetic moment. 
$c^{\dag}_{I}=(c^{\dag}_{I\uparrow -},c^{\dag}_{I\downarrow -},c^{\dag}_{I\uparrow +},c^{\dag}_{I\downarrow +})$ is an electron creation operator in the Wannier state centered at ${\bf r}_I$, and
\bea
\Sigma_i=
\begin{pmatrix}
\sigma_i &0 \\
0 & \sigma_i
\end{pmatrix}
\eea
is the spin matrix of band electrons. 

Electrons and magnetic impurities are coupled through the interaction Hamiltonian $H_J$.
We introduce the virtual crystal approximation so that we can decouple the electron system and the impurity system. We assume that magnetic fluctuations are small and neglect second and higher orders of fluctuations. The expectation value of the electron spins and the impurity spins are defined as
\be
{\bm{m}}
= \sum_{i=1}^{N} 
\frac{\langle c_i^{\dagger} {\bm\Sigma} c_i \rangle }{N} ,\hspace{5mm} {\bm M}=\sum_{I=1}^{N_S}\frac{\langle{\bm S}({\bm r}_I)\rangle}{N_{S}},
\ee
where summations are over all sites and all impurity sites, respectively. 
Finally we obtain the effective Hamiltonian decoupled into the electron part and the magnetic impurity part written as
\be
H=H_e^{{\rm MF}}+H_S^{{\rm MF}}-NJ\bm M \cdot \bm m.
\ee
$H_e^{{\rm MF}}$ ($H_S^{{\rm MF}}$) is the effective Hamiltonian of electrons (local spins),
\be
H_e^{{\rm MF}}=\sum_{\bm k}c_{\bm k}^\dagger\left[ \mathcal{H}_0(\bm k)+xJ\bm M\cdot \bm \Sigma\right]c_{\bm k},
\ee
\be
H_S^{{\rm MF}}=J\bm{m}\cdot \sum_{I=1}^{N_S} \bm S(\bm r_I),
\ee
where $x$ is the concentration of magnetic impurities.
The interaction Hamiltonian $H_J$ induces the effective Zeeman interaction to electrons, which breaks TRS. 
Since the easy axis is found to be $z$ direction determined by calculating the anisotropy energy  for Bi$_2$Se$_3$, we choose $z$ direction as a quantization axis. 
\begin{figure}[tbp]
\begin{center}
\includegraphics[width=0.5\textwidth]{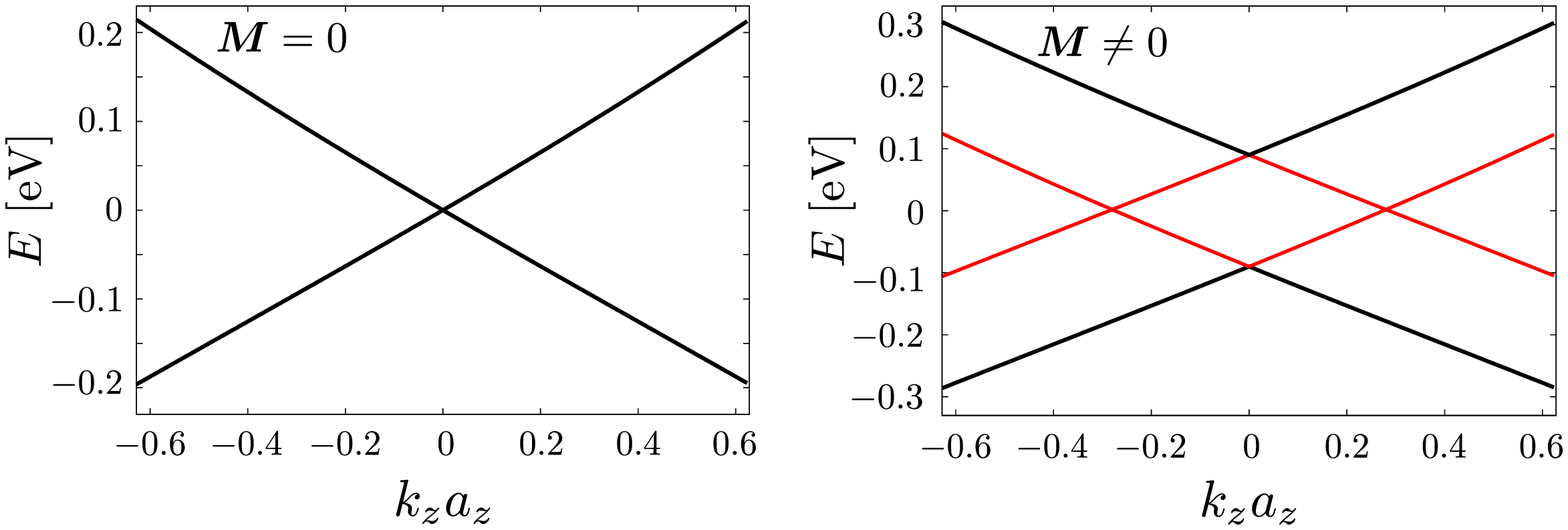}
\caption{(Color online) Magnetic impurity effect on the energy dispersion along $k_z$ axis ($k_x,k_y=0$). The parameters are $S=3/2, J=2.0$ eV, $m_0=0$ eV, and the rest of the parameters are chosen to fit the Bi$_2$Se$_3$ energy dispersion based on the first-principles calculation: $A_1=1.01$ eV, $A_2=0.32$ eV, $B_1=3.41$ eV, $B_2=0.216$ eV, $D_1=1.18$ eV, $D_2=0.024$ eV and $\epsilon_F=0$ eV.}
\label{F1}
\end{center}
\end{figure}
\begin{figure*}[t]
\begin{center}
\includegraphics[width=1.05\textwidth]{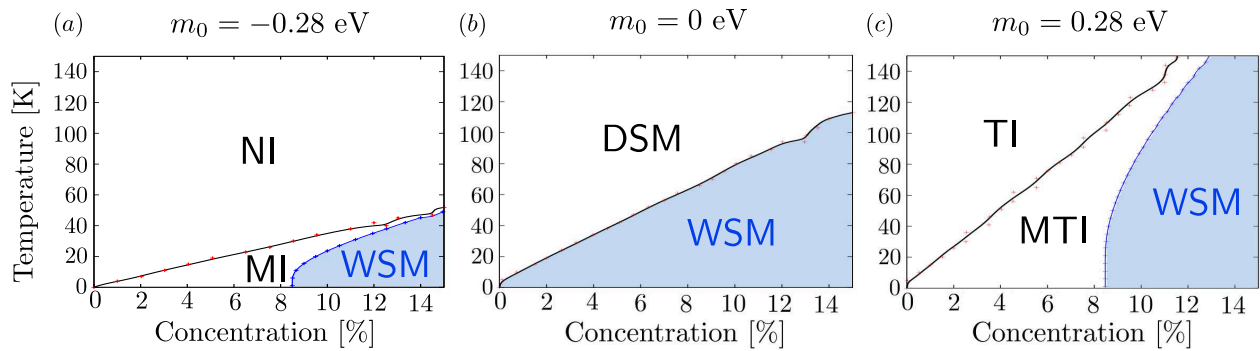}
\caption{
(Color online) The diagram of topological phases for three different strength of spin-orbit coupling, (a) $m_0=-0.28$ eV, (b) $m_0=0$ eV, and (c) $m_0=0.28$ eV.
The black line and red points correspond to the critical temperature of the magnetic transition, while the blue region describes the gapless regime whose boundary corresponds to the topological phase transition points.
In the presence of TRS, the system corresponds to (a) a normal insulator (NI), (b) a Dirac semimetal (DSM) and (c) a topological insulator (TI).
In the absence of TRS, the system becomes a magnetic insulator (MI), a magnetic topological insulator(MTI) and a Weyl semimetal (WSM) depending on the value of $m_0$ and concentration.
The black line is the magnetic phase transition temperature (Curie temperature), and the blue region corresponds to the Weyl semimetal phase with two Weyl points separating in the momentum space.

}
\label{F2}
\end{center}
\end{figure*}

The band structure of the non-magnetic phase and the ordered phase for the four-band model with $m_0=0$ eV are shown in FIG.\ref{F1}. 
In the regime without magnetic ordering (left figure in FIG.\ref{F1}), the system is called the Dirac semimetal phase which has a gap-closing point, a Dirac point which is doubly degenerated Weyl points, at the gamma point.
Despite the presence of TRS, a Dirac point is no longer protected by Kramers degeneracy and band gap may  open by perturbations, because each bands have two-fold degeneracy.
On the other hand, once the system turns into the magnetically ordered regime (right figure in FIG.\ref{F1}), effective Zeeman shift induced by the exchange interaction resolves the degeneracy of Weyl points along the symmetry broken direction, in this case, $k_z$ axis.
In the absence of TRS, it is known that there must exist same number of particles with opposite chirality, which means this state is stable as far as each Weyl points are separated in the momentum space and a band gap cannot be opened up.
In other words, the magnetic ordered regime is the Weyl semimetal phase with two topologically protected robust Weyl points.

The finite temperature magnetization is calculated by
\be
m_z=\sum_{\bm k,n}\frac{\left(U^\dagger\Sigma_z U \right)_{nn}}{N}f(\epsilon_n(\bm k)-\mu) ,
\label{eq1}
\ee
\be
M_z=-xSB_S(\beta J mS) ,
\label{eq2}
\ee
where $B_S$ is the Brillouin function 
\be
\hspace{-5mm} B_S(x)=\frac{2S+1}{2S}\mathrm{coth}\left(\frac{2S+1}{2S}x\right)-\frac{1}{2S}\mathrm{coth}\left(\frac{1}{2S}x\right).
\ee
$U$ is the matrix that diagonalizes the electron Hamiltonian $H_e^{MF}$ whose eigenvalues are given as $\varepsilon_n(\bm k)$, and $f(\epsilon-\mu)$ is the Fermi-Dirac distribution function where $\mu$ is the chemical potential.
Equations (\ref{eq1}) and (\ref{eq2}) are two coupled non-linear equations which should be solved self-consistently by an iterative procedure.\cite{Rosenberg2012}

In FIG.\ref{F2}, we show phase diagrams for the three different strength of spin-orbit coupling which are obtained by evaluating the temperature dependence of the magnetization of band electrons for the half filled system.
We choose three different $m_0$ values to be the trivial insulator, the Dirac semimetal and the topological insulator in the undoped limit.
The system, for instance, with $m_0=0$ eV and the concentration $x=5\%$ is ferromagnetically ordered up to $T_c\approx 40$K which is experimentally reachable.
This magnetic phase transition makes the band structure change as shown in FIG.\ref{F1}.
We also plot gapless regime as blue region whose boundary is considered as the transition points into the Weyl semimetal phase.
For the case $m_0=0$ eV, the magnetic phase transition corresponds to the Dirac semimetal - Weyl semimetal transition.
If the parent material is topologically non-trivial [FIG. \ref{F2} (c)], the magnetic phase transition once makes the system into the magnetic topological insulator phase\cite{QAH} whose surfaces are in the  quantized Hall regime, whereas the bulk is still insulating.
As increasing the magnetic impurity concentration, the band splitting goes larger and eventually becomes large enough to invert the conduction band and the valence band.
Even for starting from the trivial band insulator, the system possesses the ordered phase at low temperatures, and eventually turns into the Weyl semimetal phase.
Although all the systems possibly become candidate materials for the Weyl semimetal,  the Curie temperature is significantly different depending on the value of $m_0$.
\begin{figure}[b]
\includegraphics[width=0.5\textwidth]{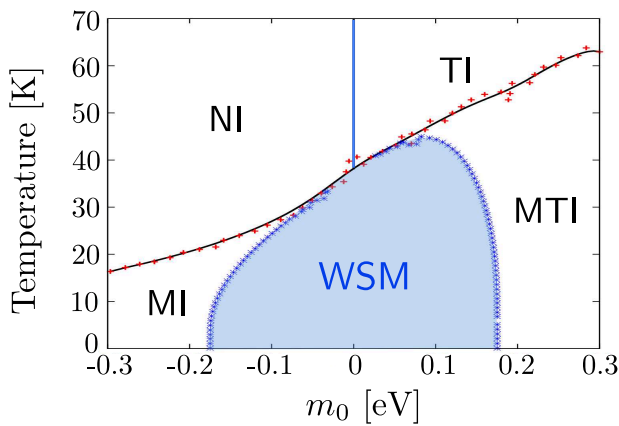}
\caption{(Color online) The diagram of topological phases as function of the strength of spin-orbit coupling and $m_0$ with magnetic impurity concentration $x=5\%$.
The black line and red points correspond to the critical temperature of the magnetic transition, while the blue region describes the gapless regime whose boundary corresponds to the topological phase transition points to the Weyl semimetal.
}
\label{F3}
\end{figure}
To evaluate the relation between the strength of spin-orbit coupling and the Curie temperature, we calculate the magnetization of band electrons with varying $m_0$.
As shown in FIG.\ref{F3}, the Curie temperature decreases as weakening the strength of the spin-orbit interaction.
In  Bi$_2$Se$_3$ family, a modification of spin-orbit coupling by varying mixing ratio of Bi$_2$Se$_3$ and Bi$_2$Te$_3$ was performed recently\cite{J.Zhang2013}.
The result, furthermore, shows that Cr-doped these materials have potential to be the stage for realizing various kinds of topological phases.
The decreasing of critical temperature indicates that the mechanism of ferromagnetism in Bi$_2$Se$_3$ family is enhanced or created by the spin-orbit interaction.
This is consistent with the fact that the ferromagnetism in those materials are mediated by the Van-Vleck paramagnetism, since the spin matrix element between a conduction band and a valence band becomes bigger with the strong spin-orbit interaction, which gives the large spin susceptibility.

Our calculation based on the mean-field approximation that the magnetic moments of dopants are replaced by their mean value.
We assumed the magnetic dopants uniformly distributed, and neglected the effect from clustering of dopants.
Our results cannot be immediately applied to the real Cr-doped Bi$_2$Se$_3$ since substituting Bi for Cr makes the strength of spin-orbit coupling weaker and the critical temperature depends on the strength of spin-orbit coupling.
Although our calculation might overestimate the critical temperature, our results can qualitatively address properties of magnetically doped Bi$_2$Se$_3$ family and the realization of the Weyl semimetal phase.

\begin{figure}[htbp]
\begin{center}
\includegraphics[width=0.45\textwidth]{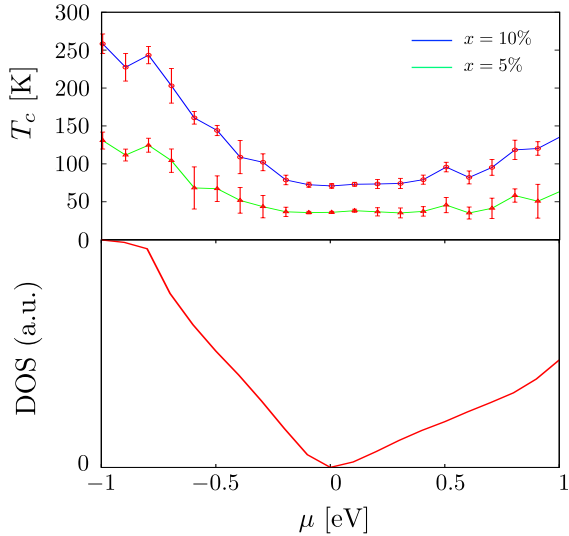}
\caption{(Color online) (Upper figure) Chemical potential dependence of Curie temperature with $m_0=0$ eV and $x=5\%,10\%$.
(Lower figure) Density of states with $m_0=0$ eV and $x=5\%$.
}
\label{F4}
\end{center}
\end{figure}

We also study the chemical potential $\mu$ dependence of the critical temperature for $m_0=0$ eV, because the chemical potential is not generally located at the Dirac point in real materials.
The result is shown in FIG. \ref{F4}.
The finite critical temperature is obtained although there is no free carriers which mediate the order in DMS systems, and the critical temperature hardly depends on the chemical potential around the Dirac point.
From these points of view, the Van-Vleck paramagnetism is dominant near the Dirac point.
On the other hand, when the chemical potential is located away form the Dirac point, we have checked that the critical temperature is proportional to the density of states (DOS).
This DOS dependence of the critical temperature is understood as Pauli paramagnetism, whose susceptibility is also proportional to the DOS, thus the Pauli paramagnetism is dominating when the carrier density is large.

\begin{figure}[tbp]
\begin{center}
\includegraphics[width=0.4\textwidth]{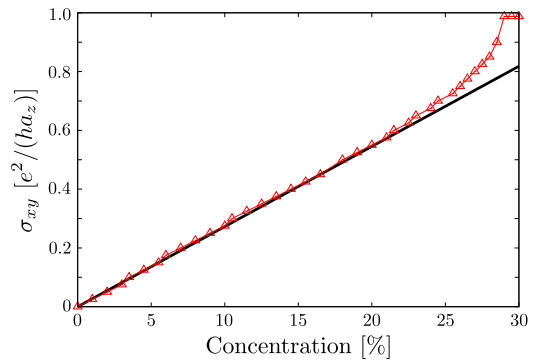}
\caption{(Color online) Anomalous Hall conductivity $\sigma_{xy}$ as a function of impurity concentration with a parameter $m_0=0$ eV at T=0K. Red triangles show the simulation result and the black solid line is the analytical result, $\sigma_{xy}=2 J M_z x / A_2a_z $.}
\label{F5}
\end{center}
\end{figure}

The Weyl semimetal phase can be characterized by the finite bulk anomalous Hall conductivity.
The anomalous Hall conductivity is calculated by considering the two-dimensional Brillouin zone which is perpendicular to the $k_z$ axis and computing the two-dimensional Chern number as a function of $k_z$, by using TKNN formula,
\be
\sigma_{xy}^{2D}(k_z)=\frac{e^2}{\hbar}\sum_{n}\int\frac{d^2k_{\perp}}{(2\pi)^2}f(\epsilon_{n\bm k_{\perp}}(k_z))\Omega_{n\bm k_\perp}^z(k_z),
\ee
\be
\Omega_{n\bm k}=\bm \nabla_{\bm k} \times \bm {A}_{n\bm k},
\ee
\be
\bm A_{n\bm{k}}=i\langle n {\bm k}| \nabla_{\bf k} |n {\bm k}\rangle
\ee
where $\Omega_{n\bm k}$ is the Berry curvature, and $\bm A_{n\bm{k}}$ is the Berry connection. 
The anomalous Hall conductivity is given by integrating the two-dimensional Chern number over $k_z$,
\be
\sigma_{xy}=\int_{-\pi/a_z}^{\pi/a_z}\frac{dk_z}{2\pi}\sigma_{xy}^{2D}(k_z)
\ee
where $a_z$ is a lattice constant.
Figure. \ref{F5} shows the numerically obtained anomalous Hall conductivity $\sigma_{xy}$ at $T=0$ K as a function of the concentration of doped magnetic impurities.
We see that the anomalous Hall conductivity is proportional to the concentration of magnetic impurities when the concentration is small.
This behavior is due to the fact that the two-dimensional Chern number is quantized in units of $e^2/h$ between two Weyl points, and vanishes at other momenta.\cite{Burkov2011}
Namely the anomalous Hall conductivity is proportional to the momentum of the Weyl point $k_D$ which is easily evaluated $k_D$ as $k_D=\pm JM_zx/A_2a_z$.
This analytical result also shows the anomalous Hall conductivity $\sigma_{xy}= \frac{JM_z}{A_2a_z}x$ which is consistent with our numerical result.
At the concentration $x\approx 30\%$, anomalous Hall conductivity reaches quantized value $\sigma_{xy}=e^2/(ha_z)$. In this region, a conduction band and a valence band are completely inverted by the strong exchange interaction and no longer possess three-dimensional massless Weyl fermions. This phase is understood as the quantum anomalous Hall insulator phase.

In conclusion, we have estimated possible conditions for realizing the Weyl semimetal phase by doping magnetic impurities in Bi$_2$Se$_3$ family within the mean-field theory.
We found that ferromagnetic ordering in these materials is seen below $T_c=40$ K with magnetic impurity concentration $x=5\%$ and $m_0=0$ eV which corresponds to the Dirac semimetal in the undoped limit.
This magnetic ordering is introduced by the Van-Vleck paramagnetism, and thus the critical temperature depends on the strength of the spin-orbit interaction.
We confirmed that the critical temperature decreases with weakening the strength of the spin-orbit interaction.
As the exchange field increases, the conduction band and the valence band invert.
The Weyl semimetal phase is realized as the gapless phase, as shown in FIG. \ref{F3}.

This work was supported by Grant-in-Aid for Scientific Research (No. 24740211 and No. 25103703) from the Ministry of Education, Culture, Sports, Science and Technology (MEXT), Japan.



\begin{thebibliography}{9}


\bibitem{Wolf2001}
S. A. Wolf, D. D. Awschalom, R. A. Buhrman, J. M. Daughton, S. von Moln\'ar, M. L. Roukes, A. Y. Chtchelkanova, and D. M. Treger, Science {\bf 294}, 1488 (2001).

\bibitem{Jungwirth2006}
T. Jungwirth, J. Ma\v{s}ek, J. KuÄ\v{c}era, and A. H. MacDonald, Rev. Mod. Phys. {\bf 78}, 809 (2006).

\bibitem{Wan2011a}
X. Wan, A. M. Turner, A. Vishwanath, and S. Y. Savrasov, Phys. Rev. B {\bf 83}, 205101 (2011).

\bibitem{Wan2011b}
X. Wan, A. Vishwanath, and S. Y. Savrasov, Phys. Rev. Lett. {\bf 108}, 146601 (2012).

\bibitem{Xu2011}
G. Xu, H. Weng, Z. Wang, X. Dai, and Z. Fang, Phys. Rev. Lett. {\bf 107}, 186806 (2011).

\bibitem{Burkov2011}
A. A. Burkov and L. Balents, Phys. Rev. Lett. {\bf 107}, 127205 (2011).

\bibitem{Weyl2014}
D. Bulmash, C. Liu, and X. Qi, Phys. Rev. B {\bf 89}, 081106 (2014).

\bibitem{TPT3}
H.-J. Kim, K.-S. Kim, J.-F. Wang, V. Kulbachinskii, K. Ogawa, M. Sasaki, A. Ohnishi, M. Kitaura, Y.-Y. Wu, L. Li, I. Yamamoto, J. Azuma, M. Kamada, and V. Dobrosavljević, Phys. Rev. Lett. {\bf 110}, 136601 (2013).

\bibitem{review_TI}
M. Z. Hasan and C. L. Kane, Rev. Mod. Phys. {\bf 82}, 3045 (2010);
X.-L. Qi and S.-C. Zhang, Rev. Mod. Phys. {\bf 83}, 1057 (2011).

\bibitem{ex_BiSe-mag}
Y. L. Chen, J.-H. Chu, J. G. Analytis, Z. K. Liu, K. Igarashi, H.-H. Kuo, X. L. Qi, S. K. Mo, R. G. Moore, D. H. Lu, M. Hashimoto, T. Sasagawa, S. C. Zhang, I. R. Fisher, Z. Hussain, and Z. X. Shen, Science {\bf 329}, 659 (2010).

\bibitem{J.Zhang2013}
J. Zhang, C.-Z. Chang, P. Tang, Z. Zhang, X. Feng, K. Li, L.-L. Wang, X. Chen, C. Liu, W. Duan, K. He, Q.-K. Xue, X. Ma, and Y. Wang, Science {\bf 339}, 1582 (2013).

\bibitem{Van-Vleck}
R. Yu, W. Zhang, H.-J. Zhang, S. Zhang, X. Dai, and Z. Fang, Science {\bf 329}, 61 (2010).

\bibitem{TPT1}
T. Sato, K. Segawa, K. Kosaka, S. Souma, K. Nakayama, K. Eto, T. Minami, Y. Ando, and T. Takahashi, Nat. Phys. {\bf 7}, 840 (2011).

\bibitem{TPT2}
H. Jin, J. Im, and A. J. Freeman, Phys. Rev. B {\bf 84}, 134408 (2011).

\bibitem{Dirac-Weyl}
H.-J. Kim, K.-S. Kim, J.-F. Wang, M. Sasaki, N. Satoh, A. Ohnishi, M. Kitaura, M. Yang, and L. Li, Phys. Rev. Lett. {\bf 111}, 246603 (2013).

\bibitem{BiSe-model}
H. Zhang, C. Liu, X. Qi, X. Dai, Z. Fang, and S. Zhang, Nat. Phys. {\bf 5}, 438 (2009).

\bibitem{BiSe-model2}
C.-X. Liu, X.-L. Qi, H. Zhang, X. Dai, Z. Fang, and S. C. Zhang, Phys. Rev. B {\bf 82}, 045122 (2010). 

\bibitem{Rosenberg2012}
G. Rosenberg and M. Franz, Phys. Rev. B {\bf 85}, 195119 (2012).

\bibitem{QAH}
C. Chang, J. Zhang, X. Feng, J. Shen, Z. Zhang, M. Guo, K. Li, Y. Ou, P. Wei, L.-L. Wang, Z.-Q. Ji, Y. Feng, S. Ji, X. Chen, J. Jia, X. Dai, Z. Fang, S.-C. Zhang, K. He, Y. Wang, L. Lu, X.-C. Ma, and Q.-K. Xue, Science {\bf 340}, 167 (2013).


\end{thebibliography}
\end{document}